\documentclass{edp-jp4}
\usepackage{graphicx}

\newcommand{\as}{\mbox{arcsec}}
\newcommand{\magarc}{\mbox{mag/arcsec$^2$}}
\begin{document}
\title{GAIA Galaxy Survey: a multi-colour galaxy survey with GAIA}
\author{Mattia Vaccari}
\address{Dipartimento di Astronomia and CISAS ``Giuseppe Colombo'',
Universit\`a di Padova, Vicolo dell'Osservatorio 2, I-35122, Padova, Italy
(\email{vaccari@pd.astro.it})}
\maketitle
\begin{abstract}
The performance expected from a galaxy survey to be carried out with GAIA,
the GAIA Galaxy Survey, is outlined.
From a statistical model of galaxy number density, size and surface brightness
distribution, and from detailed numerical simulations based on real images,
it is conservatively estimated that GAIA would be able to detect and observe
about 3 million galaxies brighter than $V \simeq 17$ and to provide
multi-colour and multi-epoch broad-band photometry of these with an
end-of-mission angular resolution of $\simeq$ 0.35 \as and a photometric
accuracy of $simeq$ 0.2 \magarc at $\mu_V = 20$ \magarc.
The substantial scientific case for performing such a survey and the
additional efforts required in terms of mission preparation, operations
and telemetry are also discussed.
\end{abstract}
%
\section{Introduction}
Since the very beginning of the feasibility studies, the GAIA mission design
was driven by the need of determining the position and the brightness of huge
numbers of stars with the uttermost accuracy.
During the Concept and Technologt Study completed in July 2000
(see \cite{GAIA2001} and \cite{GAIAsr}), however, it clearly emerged that
such a star-driven mission could also provide, several socalled by-products
which would substantially enrich its already impressive scientific yield.
In particular, it was realized that the scientific case for GAIA imaging
of high-surface brightness sky regions such as the central regions of
nearby galaxies was dramatic.

The issues connected with this opportunity have been addressed in a certain
detail in a number of studies (see \cite{GAIACUO61}, \cite{GAIACUO69},
\cite{GAIALL29} and \cite{mymasterthesis}), and such observations were
included in the mission baseline design as described in \cite{GAIAsr} under
the name of GAIA Galaxy Survey.
In this paper the current ideas on the implementation of such a survey
and on the expected performance are presented.

In Section~2 the overall strategy for galaxy detection and observation
with GAIA is outlined. In Sections~3 and~4 the detection and observation,
respectively, are discussed in greater detail and results of dedicated studies
are presented. In Section~5 the substantial scientific case for performing
such a survey is briefly sketched, and in Section~6 the additional efforts
required in terms of mission preparation, operations and telemetry are
discussed.

Only the measurement of relatively large enhancements of the surface
brightness with respect to the sky background is discussed here, although
it is believed that the rather similar issue of the observation of
Galactic Nebulae could be addressed along the same lines. The measurement
of the surface brightness of the sky background, instead, calls for a
substantially different approach, as described in \cite{GAIACUO92}.
\section{Galaxy Detection and Observation with GAIA}
During its scientific operations, the GAIA satellite will continuously spin
about its symmetry axis, and the charges contained in the CCD pixels will
correspondingly be shifted along-scan to integrate the image for a
longer exposure time, a technique known as Time-Delay Integration (TDI).
In order to limit the CCDs' reading frequency and the corresponding
readnoise and telemetry rate, a dedicated CCD readout process was
devised, consisting in detecting objects as they enter the field of view,
determining their position, magnitude and signal-to-noise ratio, and,
if the latter exceeds a certain limit, collecting data from regions
around such stars only. While this approach was the subject of detailed
studies as far as point-like objects were concerned, leading to the
definition of the Astro telescopes' focal plane and the CCD binning
strategy described e.g. in \cite{GAIACUO91}, the process of galaxy
detection and observation called for a rather different approach in
order to identify and measure faint surface brightness variations.
As first suggested in \cite{GAIACUO32}, galaxies could be detected in
the Astro Sky Mapper (ASM) as an average surface brightness significantly
in excess over the local sky background and observed in different colours in
the Broad Band Photometer (BBP). This observing strategy preventing from
optimally observing stars whenever a galaxy is being observed (which is
however less than 1\% percent of the time), it could be implemented in one
of the two Astros only.

In order to follow this general idea, one had to optimize:
\begin{itemize}
\item
the size of the areas over which the average surface brightness and the local
sky background values are computed. 
It is presently envisaged to determine the former through trimmed median
filtering of ASM\,1 samples of $2 \times 2$ pixels and the latter over a few
degrees of scan with the same algorithm.
However, the exact method used to determine the local sky background is not
of interest for our purposes, since the readnoise is by far the dominant
noise source.
\item
the value for the detection area and $S/N$ limit so that useful data are
transmitted to the ground without being swamped in less interesting data
from the Milky Way or the zodiacal light.
The larger the detection area, the fainter the detection limit can be for
objects of constant surface brightness, if the error on the sky background
is negligible. On the other hand, the detection area should be small enough
that a large number of small objects would not be missed.
\item
the sampling scheme for galaxy observations, so as to establish a trade-off
between angular resolution, readnoise and telemetry.
As with stars, a larger sample size yields a smaller error on the (average)
surface brightness, lower readnoise and telemetry rate but also a lower
angular resolution. This aspect is critical for the observations of galaxies,
since their potentially very large angular extension will require in some
cases the full readout of CCDs.
\end{itemize}

In the context of the studies carried out to demonstrate the feasibility of
the GAIA Galaxy Survey (summarized in \cite{mymasterthesis}), these
questions have been given satisfactory (if subject to further improvements)
answers.
\section{Detection via Statistical Formulae}
As for the detection, a statistical model of galaxy number density, size
and surface brightness distribution was developed in order to characterize
the ``typical'' galaxy (see \cite{GAIACUO61}).
While this model obviously cannot do justice to the well-known strong
individuality displayed by many galaxies, it is believed to yield sufficiently
reliable results when, as in our case, only statistical properties, i.e.\
properties averaged over large samples, are of interest.
On the basis of such a model, adopting current estimates of GAIA
sensitivity and noise, and adapting the statistical formulae for the
estimation of the $S/N$ obtained in the observation of point-like objects
presented in \cite{GAIACUO53} to the case of extended objects, one can
estimate whether a given galaxy could be significantly detected above the
sky background.

\begin{figure}
\centering
\includegraphics[width=\textwidth,clip=]{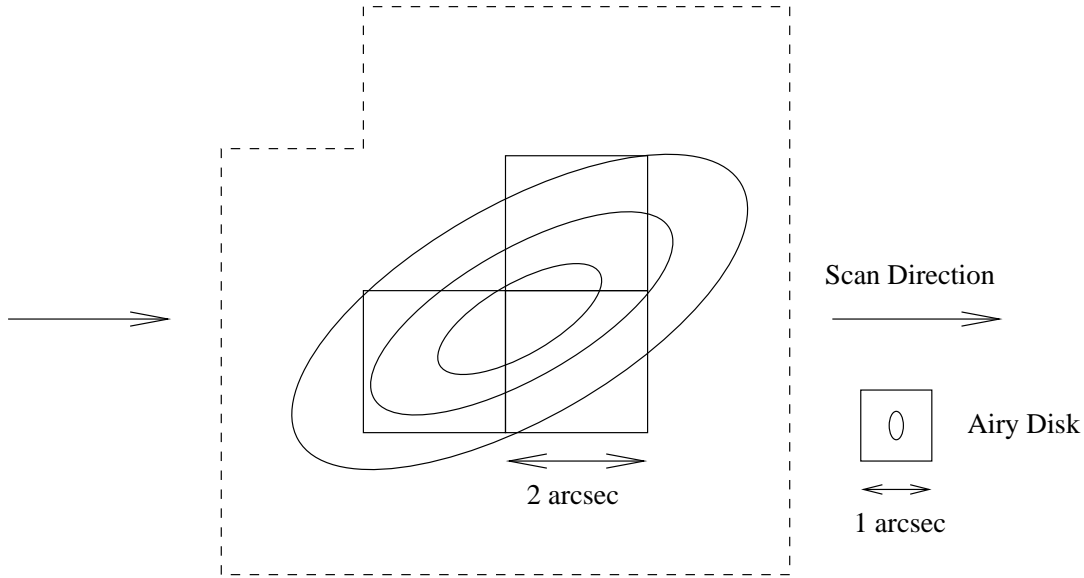}
\caption{Galaxy Detection: a galaxy is detected as an average surface
brightness significantly in excess with respect to the sky background
over areas of $2 \times 2$ \mbox{arcsec$^2$}. Data from these areas
(solid lines) and from the surrounding regions (dashed line) should then
be readout from the CCDs and transmitted to the ground.}
\label{detection.fig}
\end{figure}

It is thus concluded that a typical galaxy of $I=17$ would be detected
about 60\% of the times (i.e.\ 50 times on average during a 5-year mission)
with a $S/N > 4$ using an area of $2 \times 2$ \mbox{arcsec$^2$} for the
detection, as shown in Figure~\ref{detection.fig}.
According to our afore-mentioned statistical model, there are about
3 million galaxies brighter than this limit away from the Galactic plane
(i.e.\ with $|b|>15$, where galaxy detection shouldn't be hampered by
the high density of stars). It can thus be conservatively concluded that
GAIA would be able to reliably detect at least 3 million galaxies, in
agreement within a factor of two with the more optimistic estimation
obtained in \cite{GAIALL29} following a different approach.
Bright galaxies such as those appearing in \cite{RC3} would thus typically
be detected down to $d_{det} \simeq 400 \mathrm{Mpc}$ or $z_{det} \simeq 0.1$.
\section{Galaxy Observation via Numerical Simulations}
In order to assess the angular resolution and the photometric accuracy
obtainable from GAIA galaxy observations, complete simulation software
based on HST WFPC2 images and generating representative all-mission sets
of GAIA observations of the same field was developed.
Among other things, this has also allowed to determine the optimal sample
size and to test different stacking techniques.
The Astro PSF resulting from different smearing effects was modelled using
the tools provided in \cite{GAIALL25}, whereas the expected CCD readnoise
was calculated using the formulae given in \cite{GAIAMV04} and conservatively
assuming full CCD readout in all cases.

\begin{figure}
\centering
\includegraphics[width=0.475\textwidth,clip=]{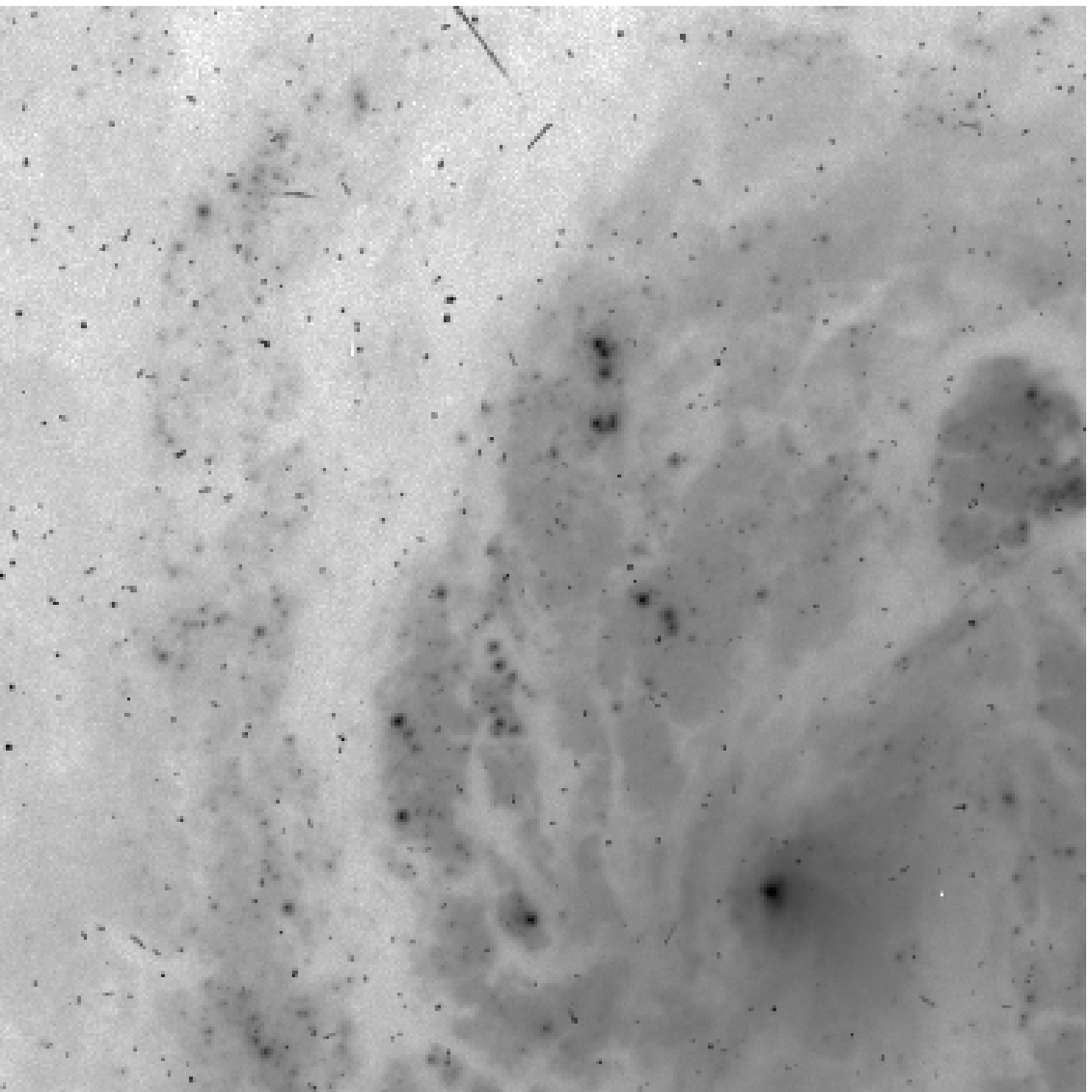}
\hfill
\includegraphics[width=0.475\textwidth,clip=]{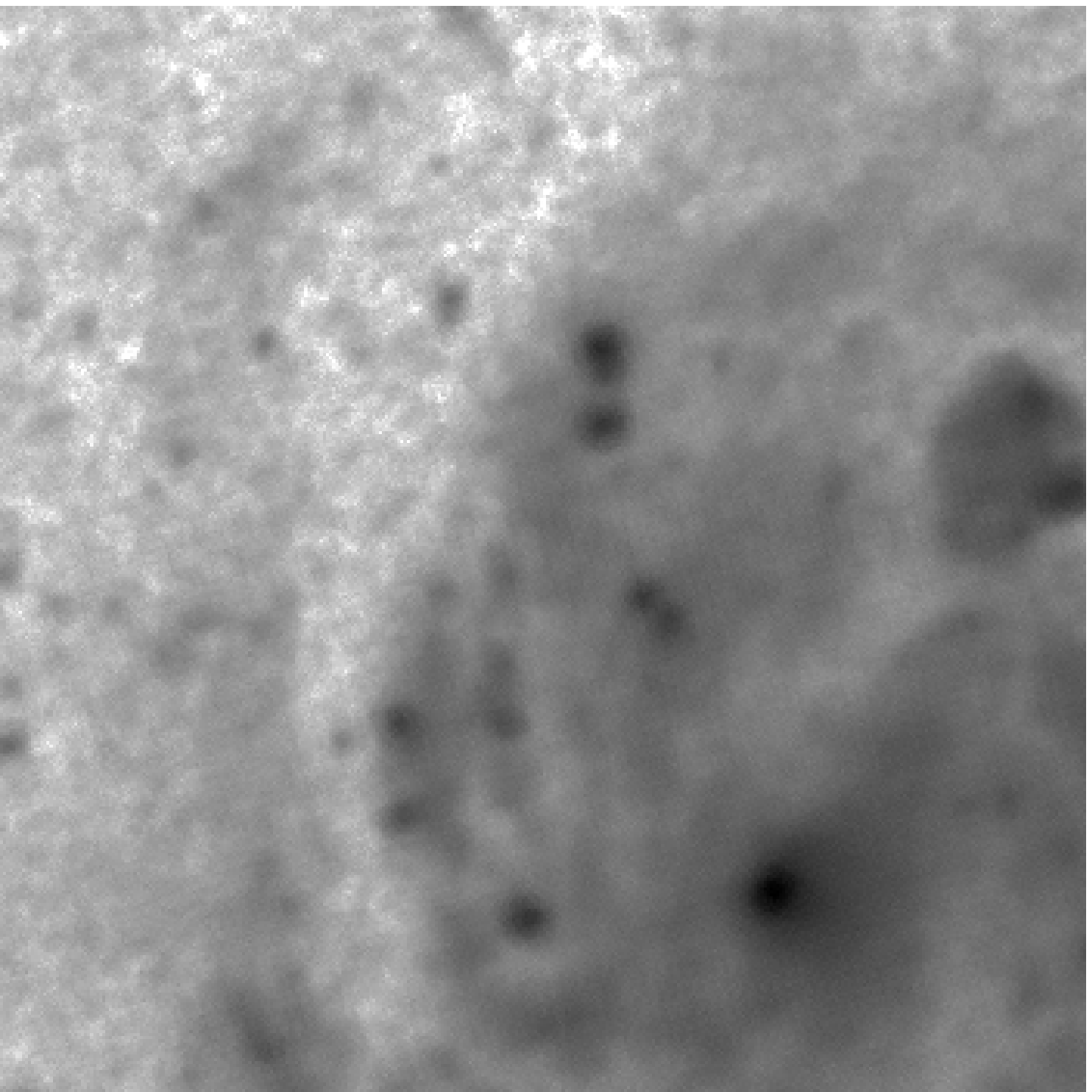}
\caption{HST Images and GAIA Flux Maps:
HST WFPC2 900~s image of the M100 Spiral Galaxy and GAIA BBP simulated flux map
obtained from 50 observations, thus with a total effective exposure time
of 45~s. The side of both images is 16 arcmin.}
\label{fluxmap.fig}
\end{figure}

The results of the simulations are shown in Figure~\ref{fluxmap.fig},
where the original HST WFPC2 900~s image is compared with GAIA BBP flux map
obtained from stacking of 50 observations of 0.9 s each, i.e.\ with an
effective exposure time of 45 s, and a sample size of $6 \times 4$ pixels,
chosen as the best trade-off between the needs of angular resolution and
photometric accuracy.
Notwithstanding the great difference in exposure time, most details are
still clearly visible in GAIA flux map, and the median photometric accuracy
is of $\simeq$ 0.2 \magarc at a median surface brightness of $\mu_V \simeq$
20 \magarc, interestingly very similar to the predictions based on statistical
formulae taking into account photon noise and readnoise.

\begin{figure}
\centering
\includegraphics[width=0.475\textwidth,clip=]{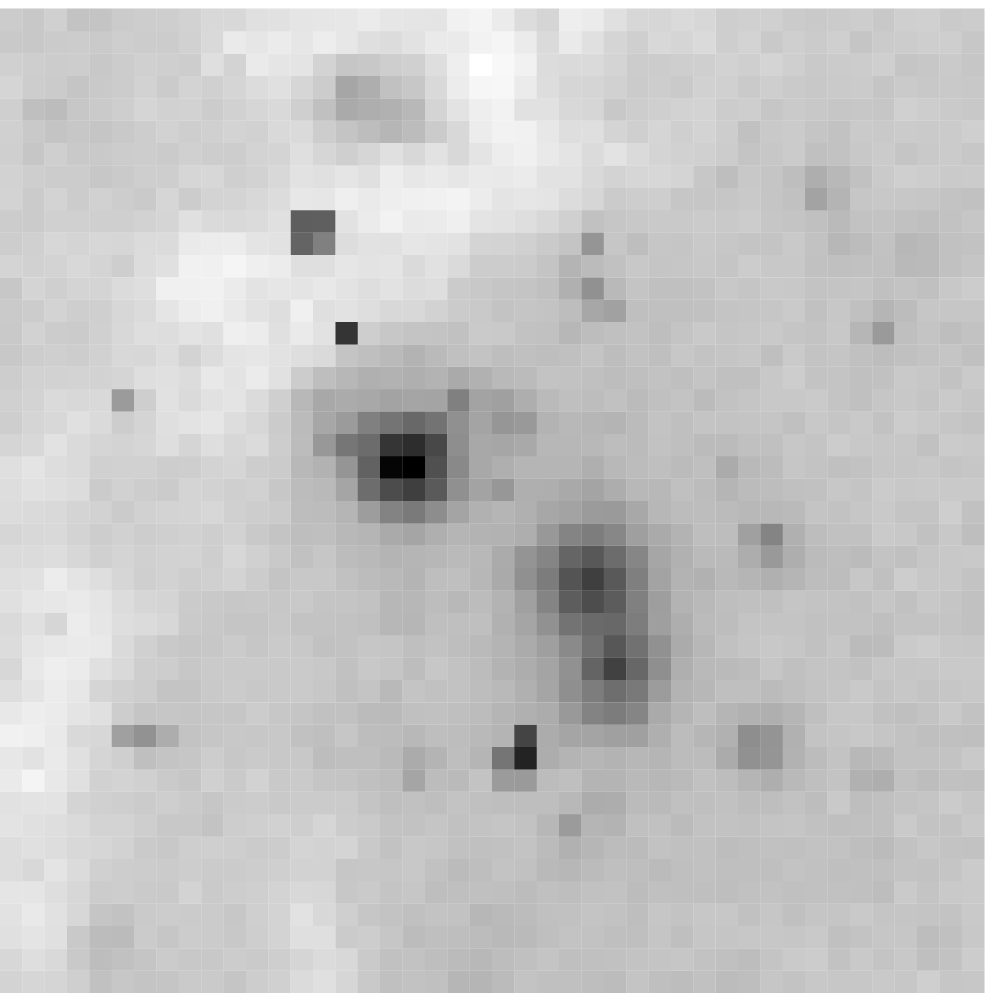}
\hfill
\includegraphics[width=0.475\textwidth,clip=]{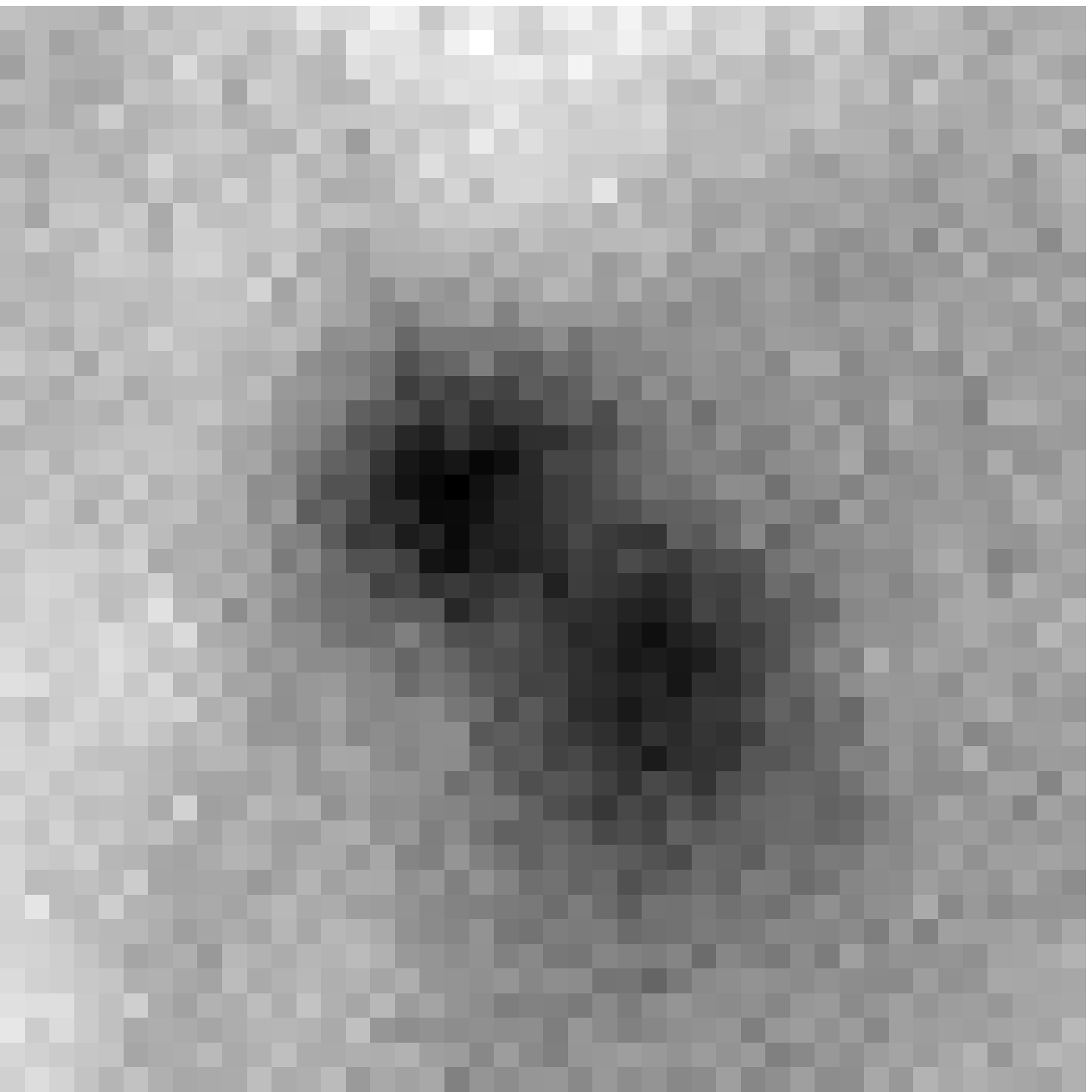}
\caption{Angular Resolution of GAIA Flux Maps: a small portion near the center
of the same HST WFPC2 image and GAIA BBP flux map shown in
Figure~\ref{fluxmap.fig}, showing two HII regions with a separation of about
\mbox{0.5~arcsec}.}
\label{fluxmap_detail.fig}
\end{figure}

The angular resolution achievable in GAIA BBP flux maps can instead be
evaluated in Figure~\ref{fluxmap_detail.fig}, where two bright HII regions
with a separation of less than 0.5 arcsec are clearly resolved in GAIA flux
map as well. More accurate estimates of the angular resolution based on model
PSFs give an angular resolution as low as 0.35 arcsec, comparable to superb
ground-based observing sites in the rare moments of excellent seeing.
Galaxy observations could therefore be profitably carried out in Astro~2,
where a sample size of $6 \times 4$ \mbox{pixels} is not in conflict with the
baseline sample size of $6 \times 8$ \mbox{pixels} adopted for the observation
of stars.
\section{Scientific Case}
The scientific case for a wide-angle high-resolution photometric (and
possibly astrometric, even though the latter issue has not been studied
in detail as yet) survey of bright galaxies is enormous.
The high reliability catalogues extending down to low Galactic latitudes
and high spatial resolution imaging of all sufficiently high surface
brightness galaxies provided by GAIA will make up a remarkably vast,
homogeneous and well-defined database that will have a tremendous impact
at two levels: for statistical analysis of the photometric structure of
the central regions of tens of thousands of well-resolved galaxies, and
for the study of the large scale structure of the Local Universe from the
spatial distribution of all detected galaxies.
Detailed analysis of the inner luminosity profiles of a large sample of
galaxies will define the true incidence of core structures and complex
morphologies. Inner color gradients will map recent star formation and dust
lanes, and central luminosity cusps may indicate massive black holes.
Multi-colour and multi-epoch information will allow the identification
of variable sources for dedicated follow-up by other telescopes.
More in general, in conjunction with the proposed high-resolution survey for
the detection of stars fainter than the nominal detection limit and the
low-resolution survey of the sky background surface brightness, GAIA
promises to yield the first uniform measurements of brightness gradients
of the ``real sky'' at all spatial scales.
\section{Mission Preparation, Operations and Telemetry}
Several aspects of the picture which was outlined require careful definition.
The galaxy detection process must be validated through the development
of a dedicated algorithm, suitable for the reliable detection of faint
extended objects on a bright complex background.
One must then make sure that the Astro CCDs can be operated so that they can
switch to the galaxy observation mode, possibly reading the full CCD, whenever
a galaxy is detected.
The opportunity of galaxy observations in the AF17 and in the SSM, where the
much higher sensitivity and spectral resolution, respectively, would result
in a much higher accuracy in surface photometry and astrophysical
characterization of the observed galaxies, should be discussed.
Finally, but perhaps most importantly, it must be evaluated how the
required telemetry (on average 120 \mbox{kbits/s/band} before compression,
where most readout samples will carry little signal and thus allow efficient
compression) can be accommodated within the overall telemetry budget.
\section{Conclusions}
The proposed GAIA Galaxy Survey will provide a nearly all-sky, multi-color and
multi-epoch astrometric and photometric galaxy survey.
In the framework of the present mission design, the feasibility, scientific
case and optimization of such a survey were discussed.
From both statistical considerations and numerical simulations it appears
that galaxies could be reliably detected in the ASM\,1 within square areas of
$2 \times 2$ \mbox{arcsec$^2$} and observed in the Astro~2 BBP with a sample
size of $6 \times 4$ \mbox{pixels}.
The first choice should yield the highest number of detected galaxies without
too may false detections, whereas the second one would provide the best
trade-off between angular resolution, readnoise and telemetry.
Under the present assumptions about the instrumental performance of the
satellite payload, and provided some effort is put into its planning
in the near future, the following measurement capabilities are expected from
a 5-year mission:

\begin{itemize}
\item
At least \mbox{3~million} galaxies brighter than $I \simeq17 $ will be
detected.
\item
All detected galaxies will be observed with a 0.35 \mbox{arcsec} angular
resolution and an all-mission accuracy in surface photometry of 0.2
\magarc at 20.0 \magarc in the $V$ band.
\item
Multi-color (in the 4--5 BBP broad bands) and multi-epoch ($\simeq 50$ epochs)
information will be available for all observed objects. White light and
medium-band photometry could as well be obtained.
\end{itemize}

These outstanding measurement capabilities will result in a unique dataset
providing state-of-the-art information on galaxy spatial distribution and
surface photometry over a well-defined sample extending down to low Galactic
latitudes, which is in turn expected to yield significant scientific results
concerning the large-scale structure of the Local Universe and the multi-color
photometric structure of galaxy innermost regions.
%
%

%

\begin{thebibliography}{99}
\bibitem{RC3}
de Vaucouleurs G., de Vaucouleurs A., Corwin H.G., Buta R.J., Paturel G. and
Fouqu\'e P. 1991, Third Reference Catalogue of Bright Galaxies, Springer
\bibitem{GAIAsr}
ESA 2000, GAIA: Composition, Formation and Evolution of the Galaxy,
Concept and Technology Study Report, ESA--SCI(2000)4
\bibitem{GAIACUO32}
H{\o}g E., Fabricius C., Knude J. and Makarov V.V. 1998,
GAIA Surveys of Nebulae and Sky Background, Technical Report SAG\_CUO\_32
\bibitem{GAIACUO53}
H{\o}g E., Fabricius C., Knude J. and Makarov V.V. 1999,
Sky Survey and Photometry by the GAIA Satellite,
Technical Report SAG\_CUO\_53 and Baltic Astronomy, 8, 25--56
\bibitem{GAIACUO91}
H{\o}g E. 2001a, Photometric and imaging performance,
Technical Report GAIA\_CUO\_91 and this volume
\bibitem{GAIACUO92}
H{\o}g E. 2001b, Multi-colour photometry with GAIA of the diffuse sky
background, Technical Report GAIA\_CUO\_92 and this volume
\bibitem{GAIALL25}
Lindegren L. 1998, Point Spread Functions for GAIA including aberrations,
Technical report GAIA\_LL\_25
\bibitem{GAIALL29}
Lindegren L. 2000, Detection of faint galaxies with GAIA,
Technical Report GAIA\_LL\_29
\bibitem{GAIA2001}
Perryman  M.A.C. et al. 2001, GAIA: Composition, Formation and Evolution of
the Galaxy, A\&A, 369, 339--363
\bibitem{GAIACUO61}
Vaccari M. and H{\o}g E. 1999a, Statistical Model of Galaxies,
Technical Report GAIA\_CUO\_61, HTML version available at
\texttt{http://hal.pd.astro.it/$\sim$\,mattia/research/smog/}
\bibitem{GAIACUO69}
Vaccari M. and H{\o}g E. 1999b, Simulated GAIA Observations of Galaxies,
Technical Report GAIA\_CUO\_69
\bibitem{mymasterthesis}
Vaccari, M. 2000, GAIA Galaxy Survey, Master Thesis, University of Padova.
Available at \texttt{http://hal.pd.astro.it/$\sim$\,mattia/research/} or
upon request writing to \texttt{vaccari@pd.astro.it}
\bibitem{GAIAMV04}
Vannier M. 1998, Noise of GAIA Astro Instrument, Technical Report GAIA\_MV\_04
\end{thebibliography}
\end{document}